\documentclass[conference,10pt]{IEEEtran}
\IEEEoverridecommandlockouts{}
\usepackage{cite}
\usepackage{amsmath,amssymb,amsfonts}
\usepackage{graphicx}
\usepackage{textcomp}
\usepackage{xcolor}

\def\BibTeX{{\rm B\kern-.05em{\sc i\kern-.025em b}\kern-.08em
    T\kern-.1667em\lower.7ex\hbox{E}\kern-.125emX}}

\usepackage{import}
\usepackage{siunitx}
\usepackage{physics}
\usepackage{bm}
\usepackage{pgf}
\usepackage{mathtools}

\usepackage{algpseudocode}
\algnewcommand\algorithmicinput{\textbf{Input:}}
\algnewcommand\Input{\item[\algorithmicinput]}

\usepackage{caption}
\usepackage{newfloat}
\DeclareFloatingEnvironment[name={Alg.}]{algorithm}

\ifCLASSOPTIONcompsoc{}
    \usepackage[caption=false,font=normalsize,labelfont=sf,textfont=sf]{subfig}
\else
    \usepackage[caption=false,font=small]{subfig}
\fi

\usepackage[capitalise]{cleveref} % has to be loaded last
\crefname{equation}{}{}
\ifCLASSOPTIONcompsoc{}
\else
    \Crefname{figure}{Fig.}{Figs.}
    \Crefname{algorithm}{Alg.}{Algs.}
    \crefname{algorithm}{Alg.}{Algs.}
\fi

\newcommand{\Real}{{\mathbb{R}}}
\newcommand{\Complex}{{\mathbb{C}}}
\newcommand{\upe}{\ensuremath{\mathrm{e}}}

\newcommand{\upj}{\ensuremath{\mathrm{j}}}
\newcommand{\uppi}{\ensuremath{\mathrm{\pi}}}
\DeclareMathOperator*{\st}{s.\!t.}
\newcommand{\inv}{{\!-1}}
\newcommand{\upT}{\ensuremath{\mathrm{T}}}
\newcommand{\upH}{\ensuremath{\mathrm{H}}}
\newcommand{\CRB}[1]{\operatorname{CRB}\!\parens{#1}}

\begin{document}

\title{Sensor Selection using the Two-Target Cramér-Rao Bound for Angle of Arrival Estimation
\thanks{The work is part of a project funded by the Netherlands Organisation for Applied Scientific Research (TNO) and the Netherlands Defence Academy (NLDA).}
}

\author{\IEEEauthorblockN{
Costas A. Kokke\IEEEauthorrefmark{3},
Mario Coutiño\IEEEauthorrefmark{1},
Laura Anitori\IEEEauthorrefmark{1},
Richard Heusdens\IEEEauthorrefmark{2},
Geert Leus\IEEEauthorrefmark{3}
}
\IEEEauthorblockA{
\IEEEauthorrefmark{1}Radar Technology, Netherlands Organisation for Applied Scientific Research,
The Hague, The Netherlands\\
\IEEEauthorrefmark{2}Netherlands Defence Academy,
Den Helder, The Netherlands\\
\IEEEauthorrefmark{3}Signal Processing Systems, Delft University of Technology,
Delft, The Netherlands
}\vspace{-2em}}

\maketitle

\begin{abstract}
  Sensor selection is a useful method to help reduce data throughput, as well as computational, power, and hardware requirements, while still maintaining acceptable performance. Although minimizing the Cramér-Rao bound has been adopted previously for sparse sensing, it did not consider multiple targets and unknown source models. We propose to tackle the sensor selection problem for angle of arrival estimation using the worst-case Cramér-Rao bound of two uncorrelated sources. We cast the problem as a convex semi-definite program and retrieve the binary selection by randomized rounding. Through numerical examples related to a linear array, we illustrate the proposed method and show that it leads to the selection of elements at the edges plus the center of the linear array.
\end{abstract}

\section{Introduction\label{sec:introduction}}

Sensor selection is the problem of choosing a subset of sensors from a full set of candidate sensors, e.g., a uniform linear array (ULA). This subset should perform better than any other subset of the same size at some task, in this case, angle of arrival (AoA) estimation.
% It can be performed both offline, in the design of the array, or online, by switching between the available elements of the array.
To perform sensor selection, a metric to evaluate the quality of selection is needed. The Cramér-Rao bound (CRB) is a natural choice~\cite{godrichSensorSelectionDistributed2012,roySparsityenforcingSensorSelection2013,tohidiSparseAntennaPulse2019}, especially for offline sensor selection, since it can be used to quantify estimation performance independent of the estimation method used. The CRB does not only depend on the selection, but also on the number of targets and their AoAs. To mitigate this, we will optimize for the worst-case two-target CRB\@.

\newcommand{\calD}{\mathcal{D}}

\begin{figure*}[!tbh]
    \centering
    \includegraphics[width=\linewidth]{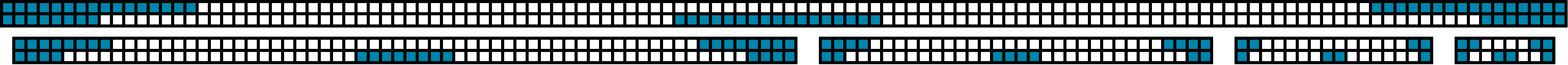}
    \caption{Array selection results for the single-target and two-target optimizations. From left-to-right then top-to-bottom, \(\qty(N, M)\) equals \(\qty(128, 32)\), \(\qty(64, 16)\), \(\qty(32, 8)\), \(\qty(16, 4)\), and \(\qty(8, 4)\), respectively.\label{fig:results1}}
\end{figure*}

\begin{figure}[!tbh]
    \centering
    \includegraphics[width=\linewidth]{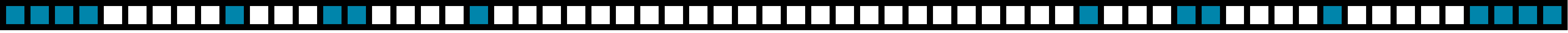}
    \caption{Array selection result for \((N, M) = (64, 16)\), when \(\calD_+\) is \num{128} equally spaced points in \(\qty[\uppi / 18, \uppi]\).\label{fig:results2}}
\end{figure}

\begin{figure}[!tb]
    \centering
    \subfloat[Varying array sizes, \(M = \frac{1}{4}N\).\label{fig:crbN}]{\input{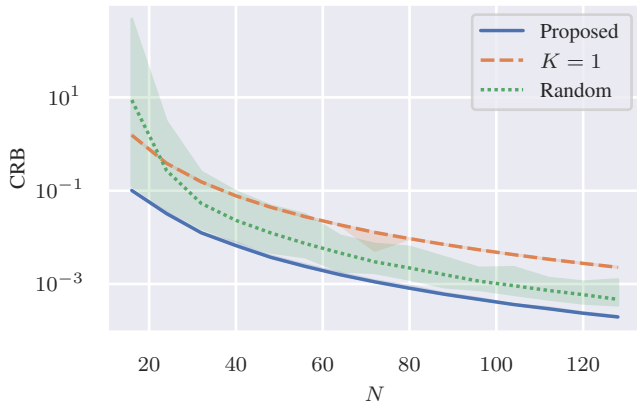}}

    \subfloat[Varying selection sizes, \(N = 128\).\label{fig:crbM}]{\input{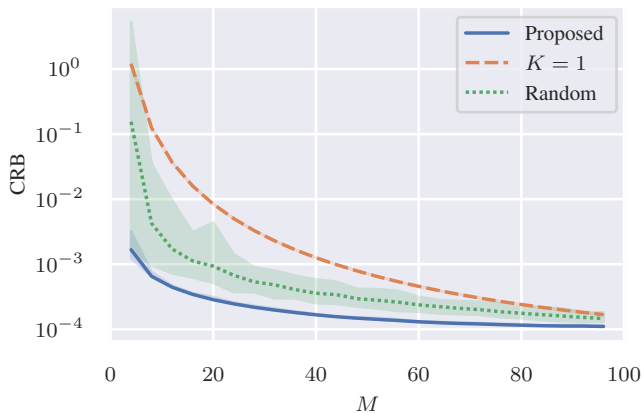}}
    \caption{Worst-case two-target CRBs. Lines indicate the average and bands the minimum and maximum values over 100 trials. In these figures \(\sigma_e^2 / \qty(2T) = 1\).\label{fig:results}}
\end{figure}

\section{Worst-Case Two-Target Cramér-Rao Bound\label{sec:crb}}

\newcommand{\Domega}{\ensuremath{\Delta\mspace{-0.5mu}\mathit{\omega}}}
\newcommand{\zo}{z\qty(\Domega)}
\newcommand{\zoc}{z^*\qty(\Domega)}
\newcommand{\bz}{\bar{z}}
\newcommand{\bzo}{\bar{z}\qty(\Domega)}
\newcommand{\bbz}{\bar{\bar{z}}}
\newcommand{\Zo}{\bm{Z}\qty(\Domega)}
\newcommand{\Z}{\bm{Z}}
\newcommand{\bzov}{\bar{\bm{z}}\qty(\Domega)}
\newcommand{\bzv}{\bar{\bm{z}}}
\renewcommand{\CRB}{\mathrm{CRB}}

Let the received data for a two-target, \(N\)-element candidate ULA over \(T\) temporal samples be given by
\begin{align*}
  \label{eq:dataModel}
  \bm X & = \bm A\qty(\bm \omega) \bm S^\upT
  + \bm E \  \in \Complex^{N \times T} \,,
\end{align*}
where \(\bm \omega \in \Real^2\) is the vector containing the target AoAs, \(\bm E\) is additive noise, and
\[
    \bm A\qty(\bm \omega) = \mqty[\bm a\qty(\omega_1) & \bm a \qty(\omega_2)] \in \Complex^{N \times K} \,,
\] is a matrix containing the array response of each target, i.e.:
\[
    \bm a \qty(\omega_k) = \mqty[1 & \upe^{\upj \omega_k} & \upe^{\upj 2 \omega_k} & \cdots & \upe^{\upj \qty(N - 1) \omega_k}]^\upT \,.
\]
\(\bm S = \mqty[\bm s_1 & \bm s_2] \in \Complex^{T \times 2} \) is a matrix containing the uncorrelated target signals.
We further assume the noise captured in \(\bm E\) to be zero-mean complex Gaussian distributed, and spatially and temporally uncorrelated with variance \(\sigma^2\).

Let \(\Domega = \omega_2 - \omega_1 \), and let \(\bm p \in \qty{0,1}^N\) be a selection vector with elements \(p_n\) where \(p_n = 1\) means the \(n\)th element of the candidate array is selected and \(p_n = 0\) means it is not.
Using the CRB as given by~\cite{stoicaMUSICMaximumLikelihood1989}, we find the CRB for both of the two uncorrelated equal-power targets to be equal to
\begin{align*}
    \CRB\qty(\Domega, \bm p) &= \frac{\sigma^2}{2T} \mathrm{Re}\qty[\bbz - \bzv^\upH\qty(\Domega) \Z^{-1}\qty(\Domega) \bzov]^{ - 1} \,,
\end{align*}
where
\begin{align*}
    \bzov &= \mqty[\sum_{n = 0}^{N - 1} p_n n \\ \sum_{n = 0}^{N - 1} p_n n \upe^{ -\upj n \Domega}] &
    \Z &= \mqty[\bm p^\upT \bm 1 & \zo \\ z^*\qty(\Domega) & \bm p^\upT \bm 1] \\
    \zo &= \sum_{n = 0}^{N - 1} p_n \upe^{\upj n \Domega} &
    \bbz &= \sum_{n = 0}^{N - 1} p_n n^2 \,.
\end{align*}

Because the CRB is equal for the two targets, we can conveniently find the worst-case for two targets as
\begin{equation}
    \label{eq:worst-crb}
    \max_{\Domega \in \mathcal{D}_+} \CRB\qty(\Domega, \bm p) \,,
\end{equation}
where the set \(\mathcal{D}_ +\) is a set of positive phase angle differences to evaluate the worst case over. Note that expressing the CRB as a function \(\Domega \) also works for non-uniform linear arrays by changing \(\zo \), \(\bzov \) and \(\bbz \) appropriately.

\section{Robust Selection using Convex Optimization\label{sec:convex}}

To find the selection \(\bm p\) that solves the minmax problem
\begin{equation*}
    \min_{\bm p}\max_{\Domega \in \mathcal{D}_+} \CRB\qty(\Domega, \bm p) \,,
\end{equation*}
we propose to turn \(\mathcal{D}_ +\) into a discrete set by uniform sampling. Then,
% by bounding the cost function and applying the Schur complement,
we find the following mixed integer semi-definite program by bounding the CRBs from above, separating the terms that depend on \(\Domega\), and applying the Shur complement:
\begin{equation*}
    \begin{aligned}
      \min_{\mathclap{\bm p, c, g}} \quad
      & c \\
      \st \quad
      & \bm 1^\upT \bm p = M \,,\ \bm p \in \qty{0, 1}^N \\
      & \mqty[\bbz - g & 1 \\ 1 & c]  \succeq 0 \\
      & \mqty[g & \bzv^\upH\qty(\Domega) \\ \bzov & \Zo] \succeq 0 \,,\ \forall \Domega \in \mathcal{D}_ +
      \,,
    \end{aligned}
\end{equation*}
where \(M\) is the desired number of selected sensors. Note that the matrix inequalities are linear in all the variables.
We solve the problem using randomized rounding~\cite{chepuriSparsityPromotingSensorSelection2015} and off-the-shelf solvers.

\section{Numerical Results and Conclusions\label{sec:results}}

To verify the proposed method and quantify its performance we have performed simulations comparing our method to random selection and selection using the single-target CRB\@. The single-target CRB optimization simply picks the \(M\) elements closest to the edges of the candidate ULA\@. The following parameters are fixed, except where noted: \num{100} randomized rounding trials, \(M = \frac{1}{4}N\), \(\calD_ +\) is \num{128} equally spaced numbers between \(1.772N^\inv \) and \SI{180}{\degree}.
\(1.772N^\inv \) is approximately the half-power beamwidth at broadside of a \(\frac{1}{2}\lambda \)-spaced ULA\@.

In \cref{fig:results1}, we vary \(N\) and \(M\) and compare the single-target and proposed optimization results to each other. Each grid is a comparison where the first row and the second row are the single-target and proposed optimization results, respectively. We see that with the proposed optimization we obtain a selection of a mix of edge and center elements. Note also that in the \(\qty(N,M) =\qty(128, 32)\) example, slightly more than half of the elements are in the center.

We have also run simulations for alternative grids, of which we show an example in \cref{fig:results2}. We again have 128 equally spaced points, but this time between \(\frac{1}{18}\pi \) and \SI{180}{\degree}. Now we see that more elements are picked more closely to the edges. When excluding larger values from \(\calD_ +\), instead of smaller ones as we have shown here, we do not observe notable changes to the results. This suggests that worst-cases occur for closely spaced targets, which makes intuitive sense.

\Cref{fig:results} summarizes the results of the more expansive simulations. The CRB that is shown in \cref{fig:results} is the worst-case two-target CRB as given by \cref{eq:worst-crb}. These results show that our optimization is successful in minimizing the worst-case. Furthermore, we can see that even a full random selection is more successful in minimizing the worst-case than using the single-target CRB as a cost function in our scenario.

We showed our method is successful in optimizing the worst-case two-target CRB for linear arrays. For future work, we plan to evaluate the robustness of our method to unknown targets by comparison to other multi-target designs. Furthermore, we plan to expand the method to planar arrays, for estimation of both azimuth and elevation AoAs.

\bibliographystyle{IEEEtran}
\bibliography{bibliography}

\end{document}